\documentclass[a4paper,twoside]{article}

\usepackage{epsfig}
\usepackage{subcaption}
\usepackage{calc}
\usepackage{amssymb}
\usepackage{amstext}
\usepackage{amsmath}
\usepackage{amsthm}
\usepackage{multicol}
\usepackage{pslatex}
\usepackage{apalike}
\usepackage[bottom]{footmisc}
\usepackage{SCITEPRESS}     
\usepackage{todonotes}
\usepackage{color}
\usepackage{colortbl}
\usepackage{float}
\restylefloat{table}
\usepackage{longtable}
\usepackage{xcolor}
\usepackage{balance}
 \usepackage{url} 

\usepackage{graphicx}
\usepackage{grffile}

\usepackage{parskip}

\usepackage{xspace}
\makeatletter
\DeclareRobustCommand\onedot{\futurelet\@let@token\@onedot}
\def\@onedot{\ifx\@let@token.\else.\null\fi\xspace}
 
\def\ie{\emph{i.e}\onedot}

\makeatother

\begin{document}

\title{Illicit Darkweb Classification via Natural-Language Processing:\\
Classifying illicit content of webpages based on textual information}

\author{\authorname{Giuseppe Cascavilla\sup{1}, Gemma Catolino\sup{2}, Mirella Sangiovanni\sup{2}}
\affiliation{\sup{1} Eindhoven University of Technology, Jheronimus Academy of Data Science, The Netherlands}
\affiliation{\sup{2}Tilburg University, Jheronimus Academy of Data Science, The Netherlands}
\email{n(ame).surname@jads.nl}
}


\abstract{
This work aims at expanding previous works done in the context of illegal activities classification, performing three different steps. First, we created a heterogeneous dataset of 113995 onion sites and dark marketplaces. Then, we compared pre-trained transferable models, i.e., ULMFit (Universal Language Model Fine-tuning), Bert (Bidirectional Encoder Representations from Transformers), and RoBERTa (Robustly optimized BERT approach) with a traditional text classification approach like LSTM (Long short-term memory) neural networks. Finally, we developed two illegal activities classification approaches, one for illicit content on the Dark Web and one for identifying the specific types of drugs. Results show that Bert obtained the best approach, classifying the dark web's general content and the types of Drugs with 96.08\% and 91.98\% of accuracy.
}

\onecolumn \maketitle \normalsize \setcounter{footnote}{0} \vfill

\section{Introduction}
The Internet, in recent times, dominates everyone's daily and professional lives, and it can be divided into three parts: The surface Web, the Deep web, and the Dark web. The surface web is the well-known part of the Internet that most of us use every day.
The Deep web is unavailable is hidden from commercial search engines, e.g., Google, since its content cannot be indexed by web crawlers. 
It is common to believe that the deep and the dark web belong to the same concept; however, the Dark web \cite{CASCAVILLA2021102258} is publicly available but can only be accessed with an encryption tool, e.g., Onion Router (Tor) \cite{Torunderattack}. Tor provides ``hidden services'' in order to host websites anonymously.

This anonymity creates the perfect environment where illegal endeavors can occur. This renders the dark web investigation extremely enticing to both law enforcement agencies and researchers to support them.
Indeed, the literature on this topic ranges from user identification \cite{AuthorshipAnalysis}, criminal motivation \cite{Criminalmotivation}, and content analysis of Tor \cite{ContentAnalysisTor} to product categorization for Darknet marketplaces \cite{Graczyk2015AutomaticPC} using natural languages processing techniques (NLP), e.g., Bert.
Nonetheless, most of the previous works \cite{Graczyk2015AutomaticPC,AuthorshipAnalysis} dealt with activities from one marketplace and did provide an extensive evaluation in terms of accuracy when comparing different text classification approaches.
This study aims to expand previous studies in three different steps. 
First, we expanded previous datasets provided in the past, expanding their size with new instances from different Darknet marketplaces. 
Then we provided a deeper comparison of pre-trained transferable models, e.g., ULMFit (Universal Language Model Fine-tuning), with a traditional text classification approach like LSTM (Long short-term memory) neural networks. Finally, we built two approaches for classifying the Dark Web's illicit activities and types of drugs.

The results show how the Bert model outperformed ULMFit and LSTM during the testing phase for both the models, i.e., illicit activities and drugs classes, while RoBERTa model obtained the lowest accuracy. According to our results, our models achieve 96.08\% accuracy for classifying illegal activities from more than one marketplace.

\section{Related Works}
\label{sec:related}

Text represents the main feature for analyzing and classifying the Darknets, thus implying the usage of text mining and Natural Language Processing techniques. For instance, Latent Dirichlet Allocation (LDA) \cite{blei2003latent} has been widely applied by the research community for identifying (i) forum discussions \cite{Tavabi2019CharacterizingAO}, or (ii) topics from Darknet \cite{DiscoveringTopicsfromDarkWebsites}. Additional studies used Bag of Words (BOW) \cite{tsai2012bag} and Term Frequency Inverse Document Frequency (TF-IDF) \cite{yun2005improved} for categorizing the content of the Darknet. For example, Nabki et al. in \cite{ClassifyingIllegal} created a dataset that encompasses the activities on the Darknet web pages. 
They found that combining TF-IDF words representation with the Logistic Regression classifier achieves 96.6\% accuracy. On the same line, Graczyk et al. in \cite{Graczyk2015AutomaticPC} attempt to provide illicit product categorization within a marketplace, using a machine learning approach. 
Choosing TF-IDF for feature extraction, the model achieved 79\% accuracy. Besides these positive results, Choi et al. \cite{terrorismrelatedarticles} showed how statistical methods like TF-IDF fall behind in text classification because they cannot comprehend the semantic meanings of the text created by people. For this reason, Sabbah et al. \cite{Hybridized} introduced a Hybridized term-weighting method to accurately identify terrorism activities with textual content from the Dark web.

\section{Methodology}
\label{sec:study}
The following subsections describe the methodology of our study. We report all the steps done and the technology used. All the experiments and data manipulation has been conducted within the Python programming environment.\\ \\
\textbf{Goal and contribution.} 
Our study aims to improve the state of the art when classifying illicit activities on the Dark web pages based on textual information.
In particular, we expanded the previous datasets provided by past studies \cite{ToRank,Graczyk2015AutomaticPC}. Then, we provide a deeper comparison of three pre-trained transferable models, e.g., ULMFit, with a traditional text classification approach like LSTM neural networks. Finally, we construct two machine learning approaches to classify the Dark Web's illicit activities and the types of drugs and analyze the performance.

\subsection{Dataset Creation}
\textbf{Data Collection.} This paper uses data from various sources from the Dark web and some from the surface web. The dataset is built to combine existing datasets (Agora, Duta10k \cite{ToRank}) and crawl the web specifically for this study. In more detail, the different data collection procedures can be divided into four parts.

The Duta10k dataset is an extension of the Duta one~\cite{ClassifyingIllegal}. 
Compared to the Duta dataset, Agora did not contain onion addresses. Hence, we considered only titles and descriptions. Nonetheless, training a text classifier on the product description is challenging since the model should be able to get textual information focused only on titles and the descriptions of the product page. 

In the context of our study, we boosted this dataset manually, extracting further pages. In particular, we collected manually several onion sites by surfing the Tor browser. As a result, we downloaded 148 onion sites. Their content ranged from counterfeit personal identification accounts (e.g., PayPal accounts), cards, and drugs and violence (weapons, hitman services). The downloaded onion pages belong to a specific onion site that cataloged all the available sites in the Tor network distributed per topic. After collecting these sites, we relied on an automatic solution for collecting HTML pages containing illicit activities from the Surface Web. The tool used to automate the collection procedure is HTTrack tool\footnote{https://www.httrack.com}. 
The copied website consists of images, links, code, and HTML pages \cite{HTTrack}.
We downloaded web sites from the Surface Web like \textit{litfakes.com} and \textit{buypurecocaineonline.com}.
Finally, we merge other data crawled by us from the Dark web, i.e., Berlusconi and the Silkroad markets.

After creating the dataset, we proceeded with the labeling phase and the data fusion process. 
\begin{itemize}
    \item Duta10k: the labeling process was already provided by the authors from \cite{ToRank}. The labels have been modified in order to improve the categories;
    \item Agora: we used the \textit{product descriptions} and \textit{titles} of the HTML pages to create new labels or to fill the labels provided by Duta10k. 
    \item Other data sources:  we manually scraped them. After extracting the text, they are hand-labeled using previous labels or creating new labels.
\end{itemize}

The Duta10k dataset initially had 26 main labels that the authors manually assigned in \cite{ToRank}. Some of them were also assigned a sub-label. A detailed description of it can be found in table \ref{table:Duta Changed} in the online appendix \cite{appendix}, specifically in the first two columns. We kept most of the original labels from the Duta10k as main classes, e.g., Services, Counterfeit Personal Identification, Counterfeit Credit-Cards, Counterfeit Money, Leaked Data, Porno, Drugs, and Violence (not included in the table for the sake of space). 
One significant modification concerned the Social Network class. In \cite{ToRank}, the authors appointed four sub-labels to it, namely, Chat, Blog, Email, and News. The Blog was altered to the main label, and the others were removed. We ignored Marketplace label due to very few occurrences.
After carefully studying each label, we merged the Hacking class with the Services one because the former referred mainly to providing hacking services. The Forum's main class was transferred from the new main classes to the Social Network main class because forums were considered a form of communication like the social networks. Therefore, having an individual class for it seemed redundant.
Similarly, we added the Human Trafficking main initial main class to the new Violence main class. Furthermore, we ignored the Cryptocurrency class as its content varied from site to site and the authors' reasons for creating this label were unclear. Lastly, we eliminated the Cryptolocker class.
All these changes to the original labels were made after considerable examination of the content of their corresponding sites.
At the end of this process, we created a list with the changed labels, and we mapped them to the initial ones. Labels like Info, eBooks, Relationships and Sex, and Drug paraphernalia and Pipes are transformed into Library Information and Drugs Paraphernalia main classes.\\

\noindent\textbf{Agora Category Labels Modifications.} The Agora web page categorization was more detailed than the Duta10k dataset. 
To adapt these categories to our work, we applied several modifications.
For the sake of space, we provide a short example of all the changes made with the labels from the Agora dataset. After carefully investigating each category and its content, all the changes were made and compared to the Duta10k main labels. We renamed the \textit{Service/Hacking} category from the Agora as \textit{Services}. Same for the category \textit{Drugs/Psychedelics/2C} that became \textit{Drugs}. \textit{Info/eBooks/Philosophy} was renamed as \textit{Library Information}. For a more detailed overview please refer to table \ref{table:Agora examples} in the online appendix \cite{appendix}.
After being transformed, the labels from Agora decreased from 104 to 15. We introduced three new labels from Agora to the final classes that did not exist before in the Duta10k, namely, Counterfeit Products, Counterfeit Coupons, and Accounts. We renamed the Agora labels to normalize the dataset with more details to the Duta10k dataset and introduce new categories.\\

\noindent\textbf{Drugs Labels.} During the labeling procedure from both Agora and Duta10k, we discovered several sub-labels describing in more detail the main labels. However, we considered only sub-classes referring to drugs for the classification task. There are two different explanations of why only drug sub-classes were considered. Firstly, using all the sub-labels implied that the classification methods should have been modified to a hierarchical format. Secondly, according to the latest studies, the interest in drug sales increases on the Dark Web. In particular, researchers \cite{TheEconomicFunctioningofOnlineDrugsMarkets}, \cite{herbalcannabis}, \cite{TorMarketplaces} and \cite{UndergroundMarketplaces} investigated  marketplaces like Silkroad for their illicit drugs content.
We created drug sub-classes from the detailed descriptions of the Agora labels. The finalized drugs sub-classes are 49, and a detailed depiction of them can be found in table \ref{table:drugs_subclass} in the online appendix \cite{appendix}. All these sub-classes correspond only to the Drugs main class.\\

\noindent \textbf{Final Classes.} As previously mentioned, the final labels are generated from blending and normalizing the Agora with the Duta10k dataset. The other data sources, particularly the self-collected pages, were manually labeled after creating the final classes. Table \ref{table:finalclass} presents the final 19 main classes. 

\begin{table}[h!]
\centering
\scalebox{0.7}{
\begin{tabular}{|cc|}
\hline
\multicolumn{2}{|c|}{Final   Main Classes}                                                                                 \\ \hline
\multicolumn{1}{|c|}{Accounts}                                                                       & Drugs               \\ \hline
\multicolumn{1}{|c|}{Counterfeit Coupons}                                                            & Drugs paraphernalia \\ \hline
\multicolumn{1}{|c|}{\begin{tabular}[c]{@{}c@{}}Counterfeit \\ Credit-Cards\end{tabular}}            & Fraud               \\ \hline
\multicolumn{1}{|c|}{Counterfeit Money}                                                              & Leaked Data         \\ \hline
\multicolumn{1}{|c|}{Counterfeit Other}                                                              & Library Information \\ \hline
\multicolumn{1}{|c|}{\begin{tabular}[c]{@{}c@{}}Counterfeit \\ Personal-Identification\end{tabular}} & Porno               \\ \hline
\multicolumn{1}{|c|}{Counterfeit Products}                                                           & Services            \\ \hline
\multicolumn{1}{|c|}{Cryptocurrency}                                                                 & Services/Money      \\ \hline
\multicolumn{1}{|c|}{Violence}                                                                       & Social Network      \\ \hline
\multicolumn{1}{|c|}{Substances for Drugs}                                                           &                     \\ \hline
\end{tabular}}
\caption{The 19 Main Classes of the Final Dataset}
\label{table:finalclass}
\end{table}

\noindent \textbf{Final Dataset.} The complete dataset contains textual information from the Duta10k dataset, the Agora, and the manually collected pages. The pages manually collected contain 640 HTML pages from CannaHome, 323 from Berlusconi market, and 1660 from Silkroad, plus other 149 manually collected pages from the Dark Web  (we used only 50 pages due to the relationship to our labels). From the pages collected using the HTTrack tool, 120 are incorporated inside the dataset from the Normal Web. The rest of the text is derived from the Agora dataset. Table \ref{table:Final_Sources} illustrates the final dataset formation.
Figure \ref{fig:Final} depicts the distribution of the classes for the whole dataset. In \cite{appendix}, figure \ref{fig:Final_nodrugs} shows the distribution without the Drugs class.

\begin{table}[h!]
\center
\scalebox{0.7}{
\begin{tabular}{|cc|}
\hline
\multicolumn{1}{|c|}{Data-source} & \textbf{Data Instances} \\ \hline
\rowcolor[HTML]{EFEFEF} 
Agora                             & 108261                                   \\
Berlouskoni                       & 323                                      \\
\rowcolor[HTML]{EFEFEF} 
CannaHome                         & 640                                      \\
Duta10k                           & 2941                                     \\
\rowcolor[HTML]{EFEFEF} 
Manual Dark Web Collection        & 50                                       \\
Normal Web                        & 120                                      \\
\rowcolor[HTML]{EFEFEF} 
SilkRoad                          & 1660                                     \\
Total of Instances Used           & 113995                                   \\ \hline
\end{tabular}}
\caption{The Data Sources in the Final Dataset}
\label{table:Final_Sources}
\end{table}



\begin{figure}[h!]
 \includegraphics[width=0.45\textwidth]{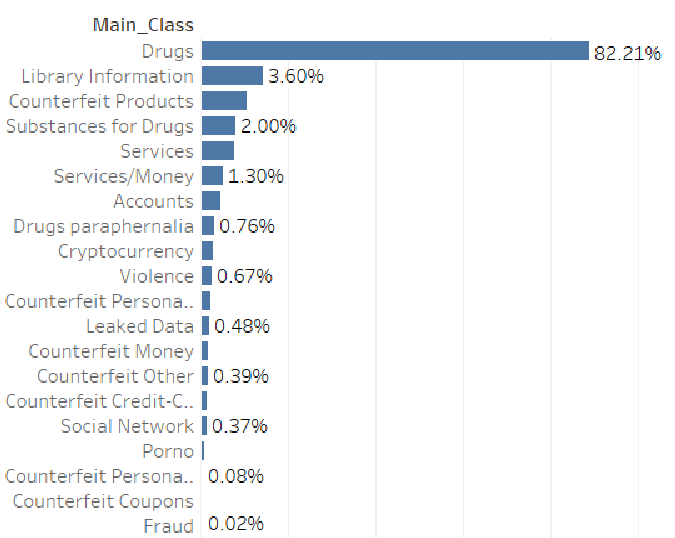}
\caption{The Distribution of the 19 Main Classes for the Final Dataset}
\label{fig:Final}
\end{figure}


\subsection{Text Preparation and Extraction}
This subsection presents all the steps followed to prepare the text for the language classification model. First, we extracted the text from the HTML pages to feed these data to the language model. 
We used Beautiful Soup\footnote{https://www.crummy.com/software/BeautifulSoup/} library to extract the text in Python. Beautiful Soup is a Python library used to survey and extract HTML content easily \cite{BS}. Also, even though all the HTML pages are stored locally, we used the module-urllib.request\footnote{https://docs.python.org/3/library/urllib.request.html} Python library to open the URLs.
Then we built two different approaches based on the market, one for random pages from the Dark web and one specific for marketplaces such as Silkroad and Berlusconi. 

In the first approach to reading the HTML files, we called the urllib.request Python library. Next, we used Beautiful Soup Python library to perform the following steps: \begin{enumerate}
    \item Eliminate all the script and style elements.
    \item Get the rest of the text from the HTML.
    \item Break the text into lines and remove the leading and trailing space on each of them.
    \item Break each of the multi-headlines into a line.
    \item Drop any blank or empty lines.
\end{enumerate} 
The second approach intends to extract only the description and the title from the HTML pages and refers only to specific marketplaces. The text from HTML from marketplaces contains repeated words (e.g., the marketplace name) and, in general, has some fixed layout. We eliminated repeated words by focusing only on the description and title. We also removed the pages that only include listings of products and do not describe a specific item. This technique was not used for all pages regardless if they belong to a marketplace because using it requires inspecting the pages, and it would not generalize well. It was utilized for marketplaces only because they have a repeated template for all the pages. Conversely, the random pages from the Dark web have a different format. Finally, after the text derived from both approaches was ready, we moved to the preparation's next part.\\

\noindent\textbf{Text Pre-processing.} Preprocessing textual data is a key component of any text classification task. The preprocessing part is usually comprised of tokenization, stop-word removal, lowercase conversion, and stemming \cite{Preprocessingimpact}. Nonetheless, the tokenization step differed from the NLP techniques used in this work. The preprocessing procedure followed in this work can be divided into six steps. 
\begin{enumerate}
\item Remove any HTML tags using Beautiful Soup Python library.
\item Remove URLs using the \textit{re} library that provides regular expression matching operations.
\item Converting contractions; i.e., when \textit{you're} becomes \textit{you are}.
\item Remove all special characters like currency signs (e.g., \$) and words that include numbers.
\item Preprocessing, to remove articles, prepositions, and pro-nouns \cite{preprocessingOverview}. Often these are not required for tasks such as sentiment analysis or text classification. Consequently, we removed stopwords using the collection from the NLTK \cite{bird2009natural} Python library.
\item In the final step, we used the WordNetLemmatizer from the NLTK library to lemmatize the processed text. Lemmatization is a way to normalize the dataset, i.e., from \textit{bought} to \textit{buy}. Lastly, all the text is transformed to lowercase.
\end{enumerate}

\subsection{Research Model}
For each of the techniques, we constructed two different models, \ie one for predicting illicit activities and one for the type of drugs. Due to the space limitation, we just reported them without details. In particular, we used Long Short Term Memory, Universal Language Model Fine-tuning, Bidirectional Encoder Representations, and Robustly optimized BERT approach.
We evaluated the performance of the models, \ie classifying illegal activities and types of drugs, through (i) accuracy and the confusion matrix, (ii) precision and recall of each model, and (iii) training time and computing power.
The transfer learning models are supposed to work with few data instances. Therefore, all four models are tested in the whole dataset (meaning the Main Classes) and a subset of the dataset (the Drugs Classes) to examine this attribute. Lastly, the procedure follows the same steps as for the classification of the types of drugs.

\section{Results}
\label{sec:results}
In this section, we report the results achieved. We discuss the performance of the models during training and the comparison between the methods regarding the testing process, for each technique used for each model, i.e., main classifier (illicit activities) and drug types.
Due to the space limitation, we report only plots of the best models, i.e., BERT. The rest is available in the online appendix \cite{appendix}.

\noindent \textbf{LSTM.}
Both models are compiled with the same model configurations meaning the loss function, optimizer, and metrics. Specifically, we chose categorical cross-entropy for the loss function, Adam as an optimizer, and accuracy as a reference metric. The models are trained for five epochs with a batch size equal to 32. Nonetheless, the results from the training process of the Main Classes LSTM model and the Drugs Classes LSTM are notably different. First of all, the computing time varies per model, as the Main Classes model needed approximately 1 hour and 24 minutes to train while the Drugs classifier took less than one hour. 


\noindent \textbf{ULMFit.}
Starting with the first cycle, the last layer of the pre-trained ULMFit is unfrozen, and the model is trained for two epochs with a learning rate equal to $1e-01$. During the first batches of data, the validation loss is much higher than the training loss; so the model struggles to learn the validation data. In the second cycle, the last two layers are unfrozen; indeed there are parts of the data batches that seem easier for the model to learn. Lastly, in the third cycle, all the layers are used to train the model for five epochs, and it learns better than before. The training time took half hour. Like the Main Classes model, the Drugs Classes model is trained during three cycles, where the last layer, the two last layers, and the whole architecture are unfrozen, respectively. We used the same epochs per cycle as ULMFit. 
Both ULMFit classifiers overfit and have greater loss values than their LSTM counterparts.



\noindent \textbf{BERT.}
The Main Classes Bert model is trained for two cycles. In the first cycle, all the layers of the architecture are frozen except for the last two. Next, the model is trained for five epochs with a 2e-4 maximum learning rate. We can notice that the model is training adequately. There is no indication of overfitting, and the loss values are below 0.4. In the second and last cycle, the last three layers are unfrozen, and it is trained for three epochs with a lower maximum learning rate, namely 2e-5. In this cycle, the classifier starts to overfit on the training data batches probably because of the layers used, and the model becomes complex. This model took approximately one and a half hours to train. Finally, Bert is trained with the same configurations as the Main Classes one in the Drugs Classes. However, in the second cycle with three unfrozen layers, this Bert model presents less overfitting than the previous one but higher loss. A possible reason is that the Drugs Classes dataset is smaller than the whole dataset used in the other Bert.
Consequently, fewer data might improve the training results. Drugs Classes took less than an hour. So far, both Bert approaches outperform the ULMFit and LSTM models in training. The loss values are low, and the training and validation data batches are more appropriately learned.\\

\noindent \textbf{RoBERTa.}
The Main Classes RoBERTa model is trained for three cycles; the model overfits on the third and last cycle instead of Bert that overfits significantly from the second cycle. The configurations for the first two cycles are the same as for the Bert models. In the third cycle, all the layers are unfrozen, and the model ends up learning better the training batches than the validation batches. RoBERTa model in the second cycle is more prone to overfitting than the corresponding Bert.
To conclude, we can indicate that the RoBERTa and Bert models for the Main and Drugs classes present superior results than the ULMFit and the LSTM.\\
\begin{figure}[h!]
\centering
\subfloat[First cycle]{\label{fig:Bert train1} {\includegraphics[width=0.4\textwidth]{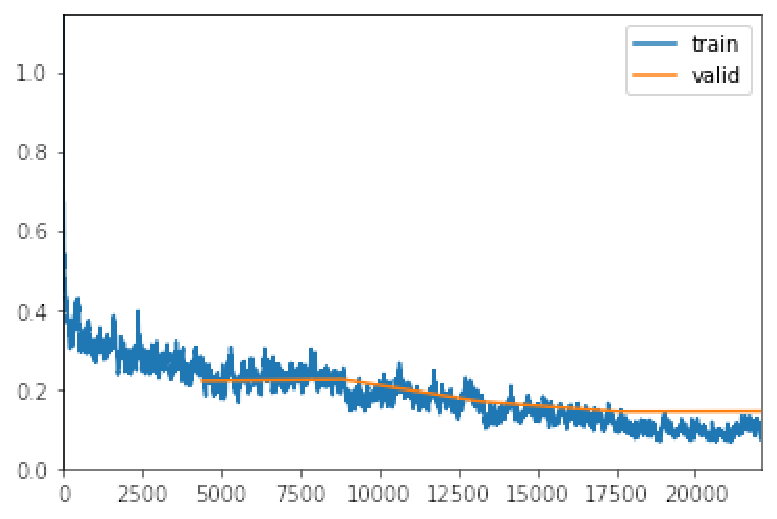}}}\hfill
\subfloat[Second cycle]{\label{fig:Bert train2} {\includegraphics[width=0.4\textwidth]{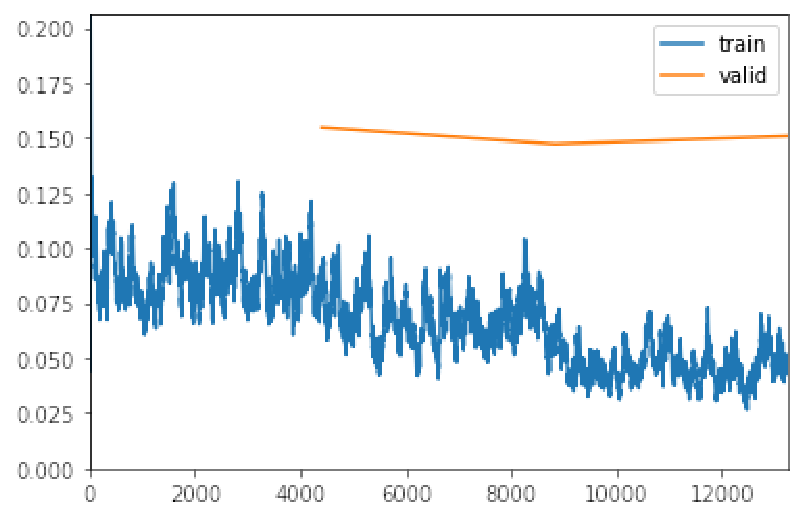}}}
\caption{Main Classes Bert model per learning cycle}
\label{fig:subfiguresBert}
\end{figure}

\begin{figure}[h!]
\centering
\subfloat[First cycle]{\label{fig:Bert_drugstrain1} {\includegraphics[width=0.4\textwidth]{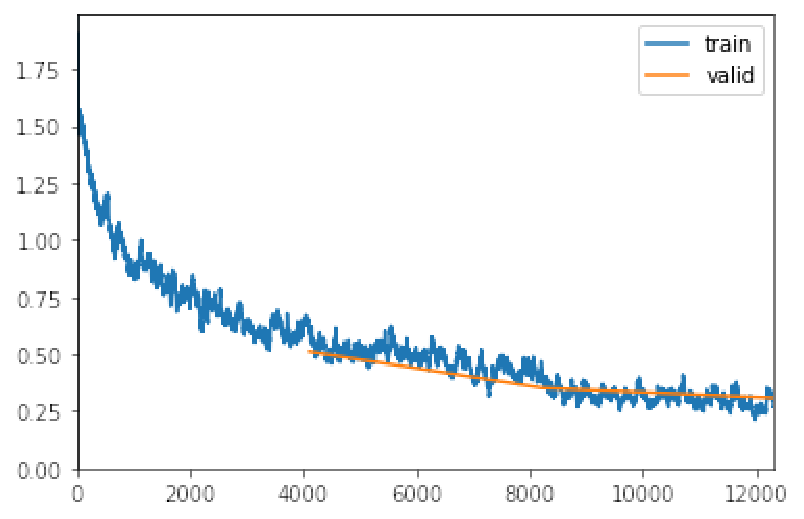}}}\hfill
\subfloat[Second cycle]{\label{fig:Bert_drugstrain2} {\includegraphics[width=0.4\textwidth]{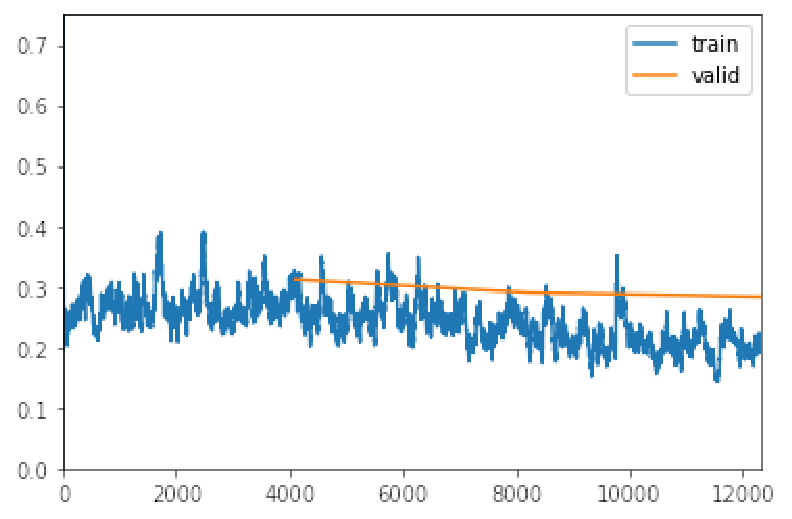}}}
\caption{Drugs Classes Bert model per learning cycle}
\label{fig:subfiguresBert_drugs}
\end{figure}


\noindent \textbf{Classification Methods Performance Comparison.}



The overall performances of the Main Classes are shown in table \ref{table:results_main}, while the ones of the Drugs in table \ref{table:results_drugs}. The tables present evaluation metrics, particularly accuracy, precision, recall, and f1 score. 

\begin{table}[h!]
\fontsize{10pt}{15}\selectfont
 \centering
\scalebox{0.55}{
\begin{tabular}{c|cccc}
\multicolumn{1}{c!{\color{black}\vrule}}{\textbf{ Main-Classes Models }} & \textbf{ Accuracy } & \textbf{ Precision } & \textbf{ Recall } & \textbf{ F1 score }  \\ 
\hline
\textbf{ Bert }                                                          & \cellcolor[HTML]{EFEFEF} \textbf{ 96.08\% }  & 0.82                 & 0.78              & 0.80                 \\
\textbf{ RoBERTa }                                                       & 95.78\%             & 0.85                 & 0.82              & 0.84                 \\
\textbf{ ULMFit }                                                        & 95.98\%             & 0.77                 & 0.74              & 0.74                 \\
\textbf{ LSTM }                                                          & 95.91\%             & 0.80                 & 0.77              & 0.78                
\end{tabular}}
\arrayrulecolor{black}
\caption{Results metrics for the Main Classes Models}
\label{table:results_main}
\end{table}

\begin{table}[h]
\fontsize{10pt}{15}\selectfont
 \centering
\scalebox{0.55}{
\begin{tabular}{c|cccc}
\multicolumn{1}{c!{\color{black}\vrule}}{\textbf{ Drugs-Classes Models }} & \textbf{ Accuracy } & \textbf{ Precision } & \textbf{ Recall } & \textbf{ F1 score }  \\ 
\hline
\textbf{ Bert-Drugs }                                                     & \cellcolor[HTML]{EFEFEF} \textbf{ 91.98\% }  & 0.85                 & 0.85              & 0.84                 \\
\textbf{ RoBERTa-Drugs }                                                  & 91.38               & 0.83                 & 0.79              & 0.80                 \\
\textbf{ ULMFit-Drugs }                                                   & 61.01               & 0.23                 & 0.19              & 0.18                 \\
\textbf{ LSTM-Drugs }                                                     & 90.58               & 0.84                 & 0.82              & 0.82                
\end{tabular}}
\arrayrulecolor{black}
\caption{Results metrics for the Drugs Classes Models}
\label{table:results_drugs}
\end{table}

Moving to the Drugs Classes models results in table \ref{table:results_drugs}, Bert has the highest accuracy score. However, the metrics scores for these models are more diverse than the previously discussed. Specifically, the ULMFit model severely underperforms compared to the rest. The Bert surpasses all three models in the metrics scores, but more importantly, the precision and recall values are the same. This is an ideal trade-off situation as the precision and recall metrics should be as close as possible. Bert is the best from the Drugs Classes models, and RoBERTa comes second. The results of the Drugs Classes models can be used to evaluate their scalability. The scalability of the models refers to how well they perform with fewer data instances. Since the Drugs Classes ULMFit model performs poorly compared to the Main Classes one, it is clear that the ULMFit approach is not scalable. The other three models have produced similarly adequate results with the whole dataset and a subset of the datasets, specifically, the drugs-related HTML pages.

Based on the results, we investigated the confusion matrix of Bert and RoBERTa approaches for both models. Unfortunately, due to the space limitation, we did not add figures, but we uploaded them in the online appendix \cite{appendix}.
When investigating the Main Classes Bert matrix, we noticed that the model mostly confuses the label Drugs with the label Substances for Drugs. In particular, when the label should have been Substances for Drugs, Bert assigns the label Drugs 26 times and vice versa. 
Regarding the RoBERTa model, the confusion matrix indicates that the Drugs and Substances for Drugs labels are the most confused, as in Bert. However, in RoBERTa, the Drugs and Substances for Drugs classes are wrongly identified 67 times. Therefore, the ability of Bert to identify the correct labels seems better.




\section{Conclusion}
\label{sec:conclusion}
This study provided several different approaches for classifying the Dark Web content that achieved good results. Specifically, the best model classified the Dark Web's general content with 96.08\% accuracy and the specific types of drugs with 91.98\% accuracy. Future studies include exploiting the hierarchical classification model and developing a similar classifier for multiple languages.

\section*{\uppercase{Acknowledgements}}
The authors thank MSc student Theodora Tzagkaraki for her valuable job, and Prof. W.J. van den Heuvel and Prof. D.A. Tamburri for providing feedback to improve the quality of the paper.

\bibliographystyle{apalike}
\bibliography{reference.bib}



\begin{table*}[h!]
\centering
\scalebox{0.8}{
\begin{tabular}{|l|l|l|} 
\hline
\textbf{ Main Class Duta10k }       & \textbf{ Sub-Class Duta10k  }                                                                              & \textbf{ New Main Class }            \\ 
\hline
Art                                 & Music                                                                                                      & X                                    \\ 
\hline
Casino                              & Gambling                                                                                                   & X                                    \\ 
\hline
\begin{tabular}[c]{@{}l@{}} Counterfeit\\Personal-Identification \end{tabular} & \begin{tabular}[c]{@{}l@{}} Driving-License\\Passport\\ID \end{tabular}                                    & \begin{tabular}[c]{@{}l@{}} Counterfeit\\ Personal-Identification\end{tabular}  \\ 
\hline
Drugs                               & \begin{tabular}[c]{@{}l@{}} Legal\\Illegal \end{tabular}                                                   & Drugs                                \\ 
\hline
Forum                               & ~                                                                                                          & Social Network                       \\ 
\hline
Hacking                             & ~                                                                                                          & Services                             \\ 
\hline
Hosting                             & \begin{tabular}[c]{@{}l@{}} Folders Directory\\Server\\Search-Engine\\Software\\File-sharing \end{tabular} & ~X
  ~                        \\ 
\hline
Human-Trafficking                   & ~                                                                                                          & Violence                             \\ 
\hline
Library                             & Books                                                                                                      & Library Information                  \\ 
\hline
Marketplace                         & \begin{tabular}[c]{@{}l@{}}    Legal \\~Illegal     \end{tabular}                                          & ~X~ ~                                \\ 
\hline
Pornography                         & \begin{tabular}[c]{@{}l@{}} Child-pornography\\~General-pornography \end{tabular}                          & Porno                                    \\ 
\hline
Social-Network                      & \begin{tabular}[c]{@{}l@{}}    Chat \\Email\\Blog\\News     \end{tabular}                                  & Social Network                       \\ 
\hline
Violence                            & \begin{tabular}[c]{@{}l@{}} Hate~\\Weapons\\Hitman \end{tabular}                                           & Violence                             \\
\hline
\end{tabular}}
\caption{The changed labels from the original Duta10k to the new main classes}
\label{table:Duta Changed}
\end{table*}

\begin{table*}[h!]
\centering
\scalebox{0.9}{
\begin{tabular}{|l|l|}
\hline
\textbf{Agora Category}      & \textbf{New Main Class} \\ \hline
Services/Hacking             & Services                \\ \hline
Data/Software                & Services                \\ \hline
Data/Accounts                & Accounts                \\ \hline
Counterfeits/Money           & Counterfeit Money       \\ \hline
Electronics                  & Counterfeit Products    \\ \hline
Data/Pirated                 & Leaked Data             \\ \hline
Jewelry                      & Counterfeit Products    \\ \hline
Counterfeits/Accessories     & Counterfeit Products    \\ \hline
Counterfeits/Watches         & Counterfeit Products    \\ \hline
Info/eBooks/Anonymity        & Library Information     \\ \hline
Counterfeits/Electronics     & Counterfeit Products    \\ \hline
Services/Travel              & Services                \\ \hline
Drugs/RCs                    & Substances for Drugs    \\ \hline
Drugs/Psychedelics/2C        & Drugs                   \\ \hline
Drugs/Opioids/Heroin         & Drugs                   \\ \hline
Drugs/Opioids/Fentanyl       & Drugs                   \\ \hline
Drugs/Opioids/Oxycodone      & Drugs                   \\ \hline
Tobacco/Paraphernalia        & Drugs paraphernalia     \\ \hline
Weapons/Ammunition           & Violence                \\ \hline
Info/eBooks/Philosophy       & Library Information     \\ \hline
Drug paraphernalia/Paper     & Drugs paraphernalia     \\ \hline
Drugs/Opioids/Dihydrocodeine & Drugs                   \\ \hline
Drugs/Dissociatives/GBL      & Drugs                   \\ \hline
Drugs/Psychedelics/Salvia    & Drugs                   \\ \hline
Drugs/Barbiturates           & Drugs                   \\ \hline
Weapons/Fireworks            & Violence                \\ \hline
Drug paraphernalia/Scales    & Drugs paraphernalia     \\ \hline
Chemicals                    & Substances for Drugs    \\ \hline
\end{tabular}}
\caption{A sample of the Agora Categories and their modification to the new Main Classes}
\label{table:Agora examples}
\end{table*}

\begin{table*}[h!]
\centering
\scalebox{0.8}{
\begin{tabular}{|l|l|} 
\hline
\multicolumn{2}{|l|}{\textbf{~ ~ ~ ~ ~ ~ ~ ~ ~ ~ ~ ~ Drugs Sub-Classes }}  \\ 
\hline
1.~~~~~ Tobacco Smoked        & 26.~~ Ecstasy MDA                               \\ 
\hline
2.~~~~~ Weight loss           & 27.~~ Psychedelics                              \\ 
\hline
3.~~~~~ Steroids              & 28.~~ Psychedelics Mescaline                    \\ 
\hline
4.~~~~~ Prescription          & 29.~~ Psychedelics 2C                           \\ 
\hline
5.~~~~~ Other                 & 30.~~ Opioids                                   \\ 
\hline
6.~~~~~ Cannabis/Weed         & 31.~~ Opioids Morphine                          \\ 
\hline
7.~~~~~ Benzos                & 32.~~ Opioids Heroin                            \\ 
\hline
8.~~~~~ Cannabis              & 33.~~ Opioids Fentanyl                          \\ 
\hline
9.~~~~~ Cannabis Concentrates & 34.~~ Opioids Oxycodone                         \\ 
\hline
10.~~ Cannabis Seeds          & 35.~~ Psychedelics 5-MeO                        \\ 
\hline
11.~~ Cannabis Hash           & 36.~~ Opioids Buprenorphine                     \\ 
\hline
12.~~ Dissociatives MXE       & 37.~~ Psychedelics Spores                       \\ 
\hline
13.~~ Dissociatives Ketamine  & 38.~~ Opioids Hydrocodone                       \\ 
\hline
14.~~ Cannabis Edibles        & 39.~~ Dissociatives                             \\ 
\hline
15.~~ Ecstasy Pills           & 40.~~ Opioids Codeine                           \\ 
\hline
16.~~ Ecstasy                 & 41.~~ Opioids Dihydrocodeine                    \\ 
\hline
17.~~ Stimulants Meth         & 42.~~ Stimulants Mephedrone                     \\ 
\hline
18.~~ Psychedelics LSD        & 43.~~ Dissociatives GBL                         \\ 
\hline
19.~~ Stimulants Speed        & 44.~~ Opioids Opium                             \\ 
\hline
20.~~ Dissociatives GHB       & 45.~~ Psychedelics Salvia                       \\ 
\hline
21.~~ Stimulants Cocaine      & 46.~~ Barbiturates                              \\ 
\hline
22.~~ Psychedelics NB         & 47.~~ Dissociatives PCP                         \\ 
\hline
23.~~ Psychedelics Mushrooms  & 48.~~ Ketamine                                  \\ 
\hline
24.~~ Ecstasy MDMA            & 49.~~ Psychedelics                              \\ 
\hline
25.~~
  Psychedelics
  DMT    & ~                                               \\
\hline
\end{tabular}}
\caption{The 49 Drugs Sub-Classes}
\label{table:drugs_subclass}
\end{table*}

\begin{figure*}[ht]
 \includegraphics[width=1\textwidth]{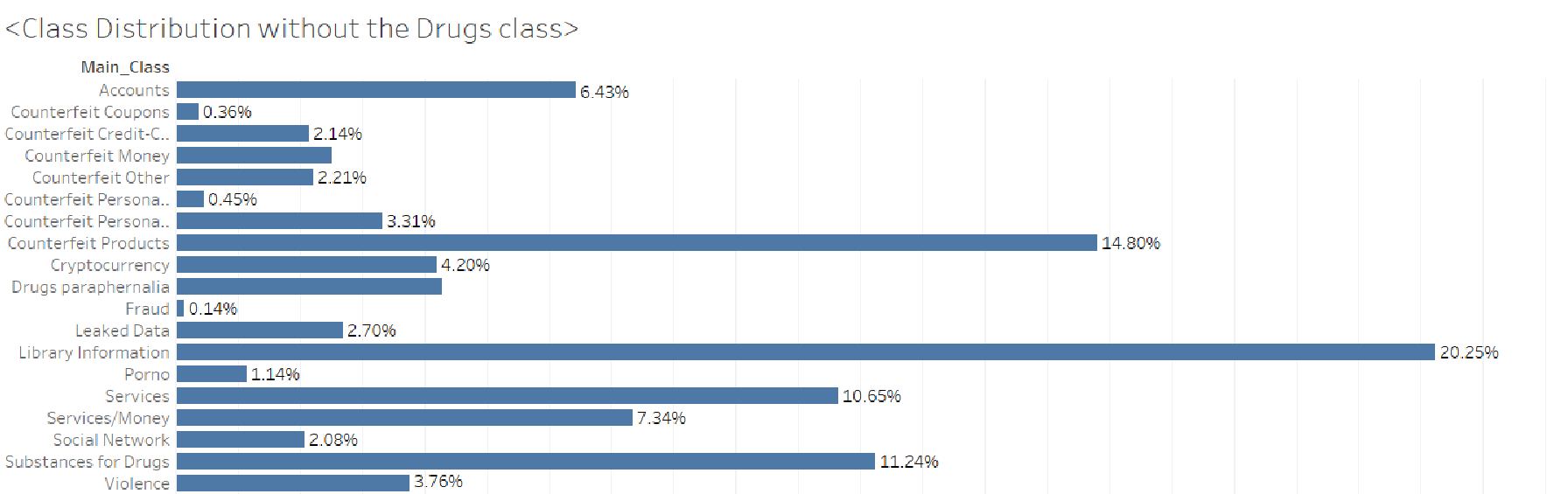}
\caption{The Distribution of the Main Classes Without the Drugs Class for the Final Dataset.}
\label{fig:Final_nodrugs}
\end{figure*}


\begin{figure*}[ht]
\subfloat[Confusion Matrix of Main Classes Bert]{\label{fig:Bertconf} {\includegraphics[width=0.5\textwidth]{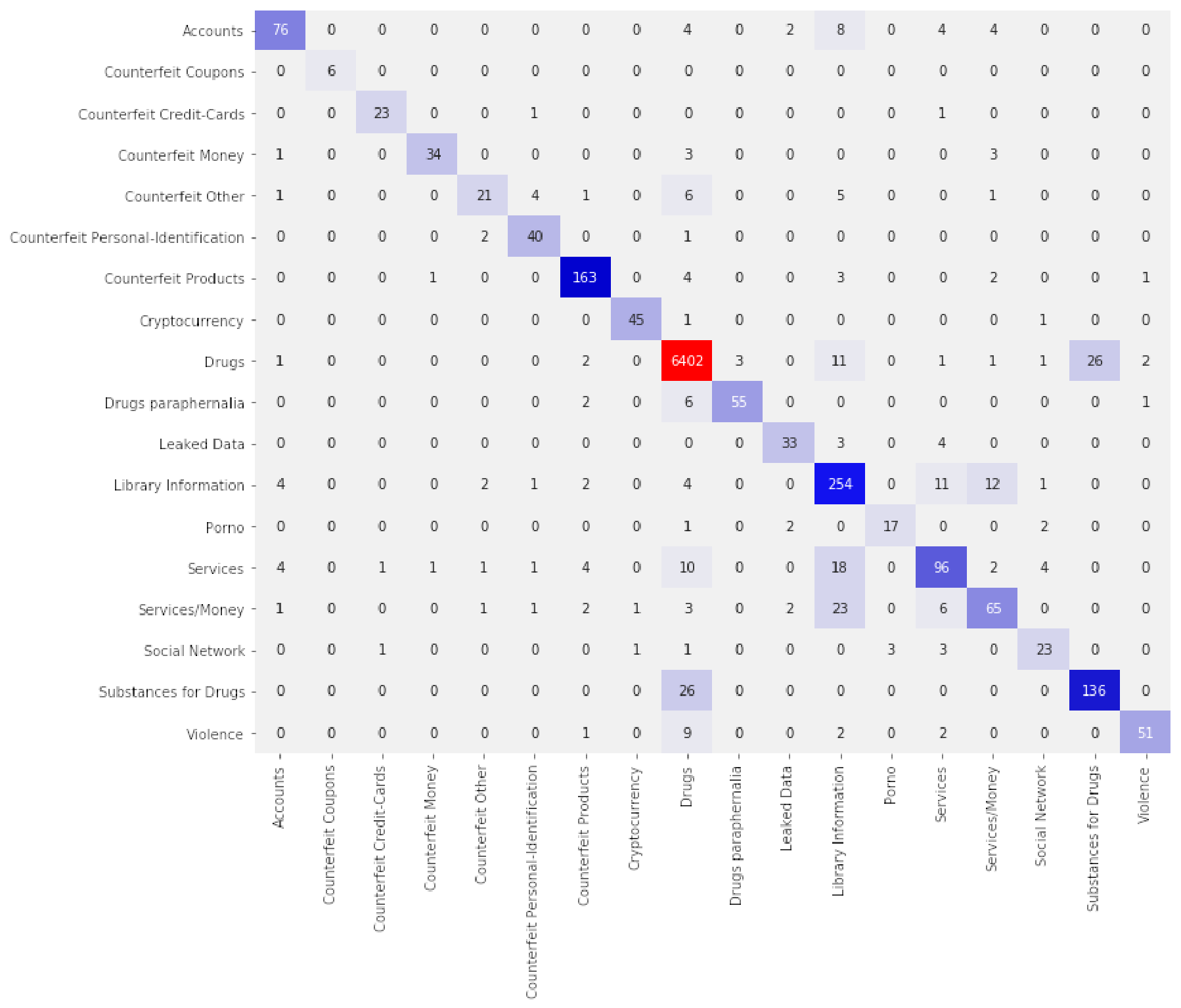}}}\hfill
\subfloat[Confusion Matrix of Main Classe RoBERTa]{\label{fig:RoBertaconf} {\includegraphics[width=0.5\textwidth]{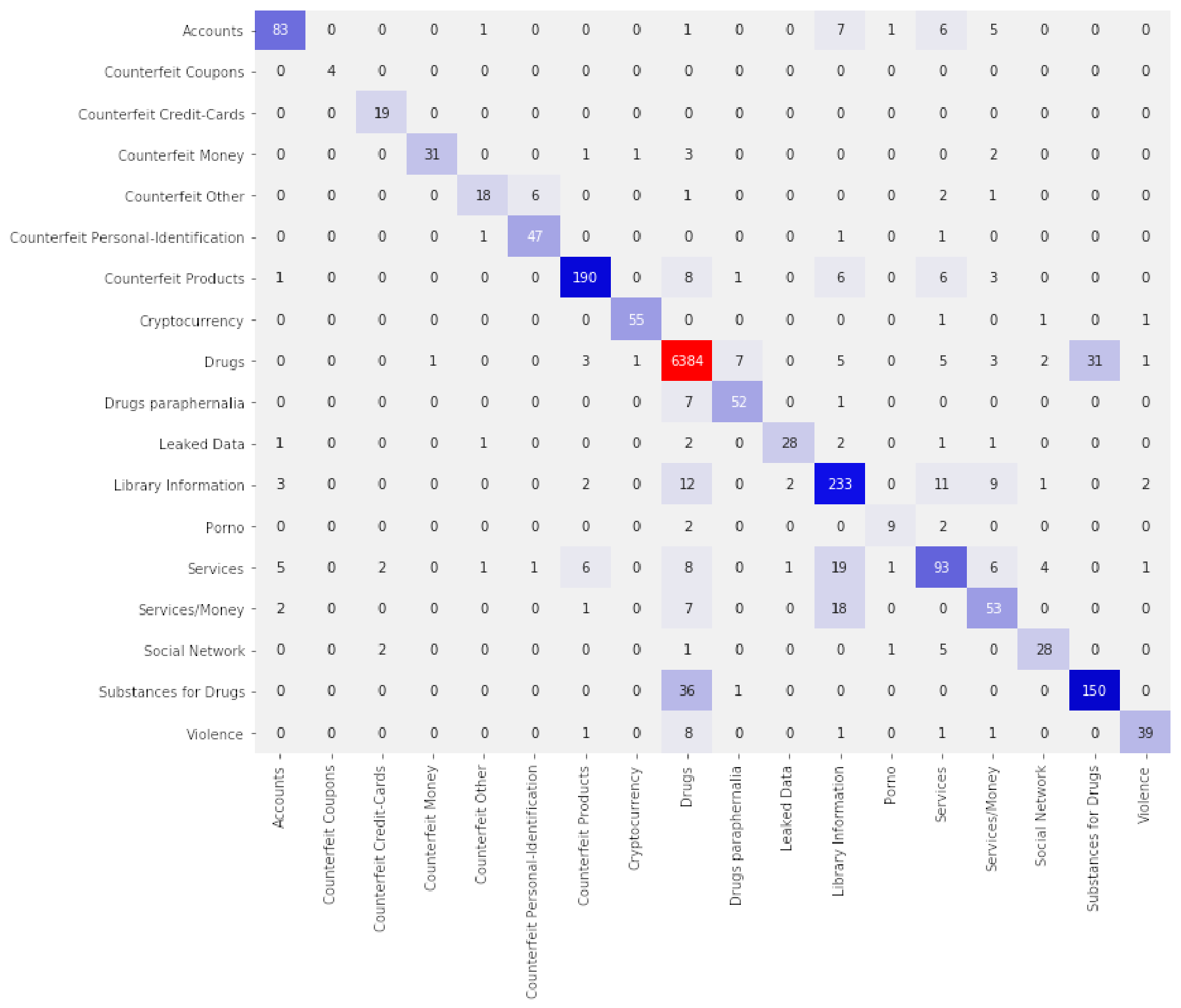}}}\hfill
\subfloat[Confusion Matrix of Drugs Classes Bert]{\label{fig:Bert_drugsconf} {\includegraphics[width=0.5\textwidth]{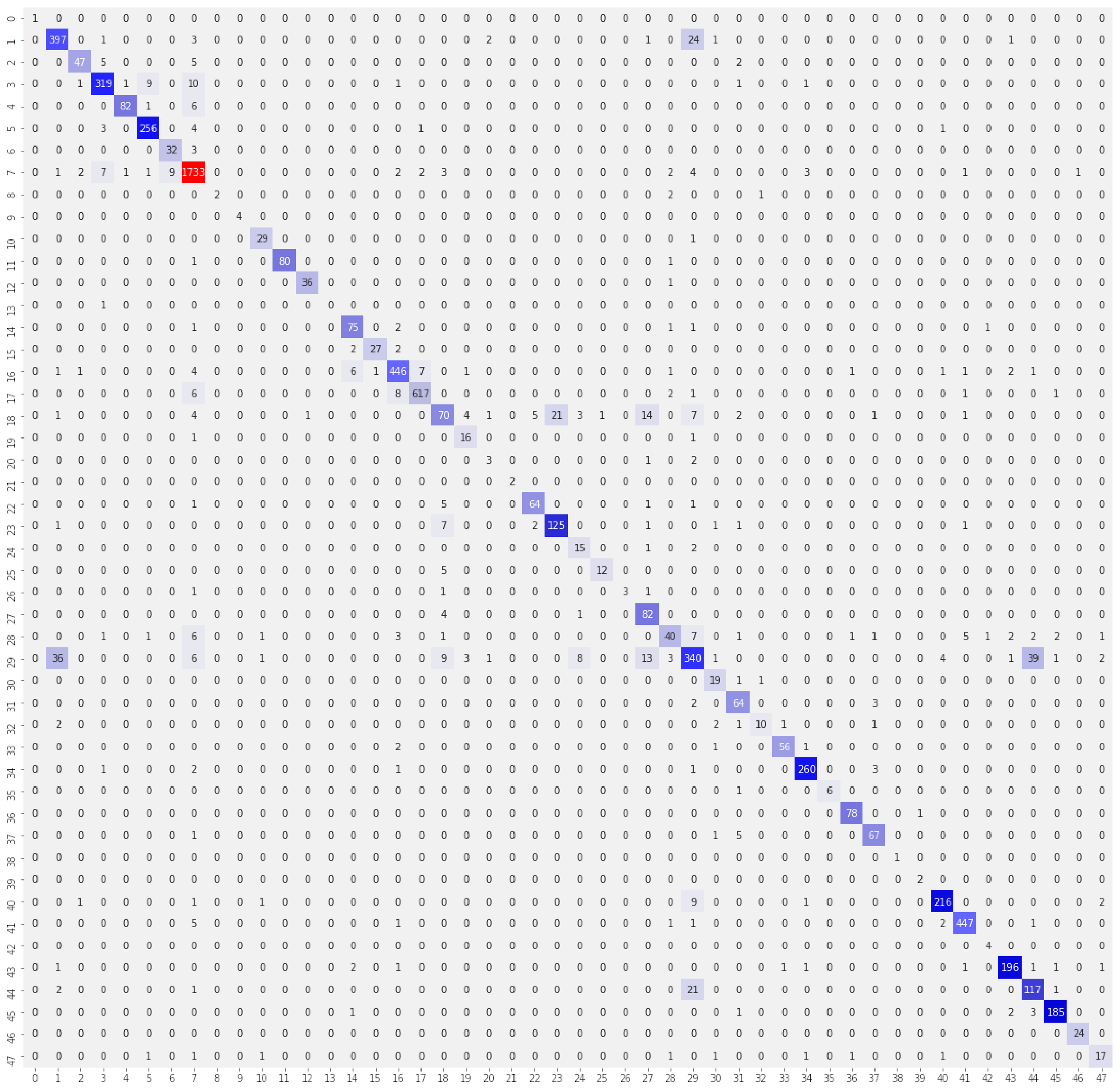}}}\hfill
\subfloat[Confusion Matrix of Drugs Classes RoBERTa]{\label{fig:Roberta_drugsconf} {\includegraphics[width=0.5\textwidth]{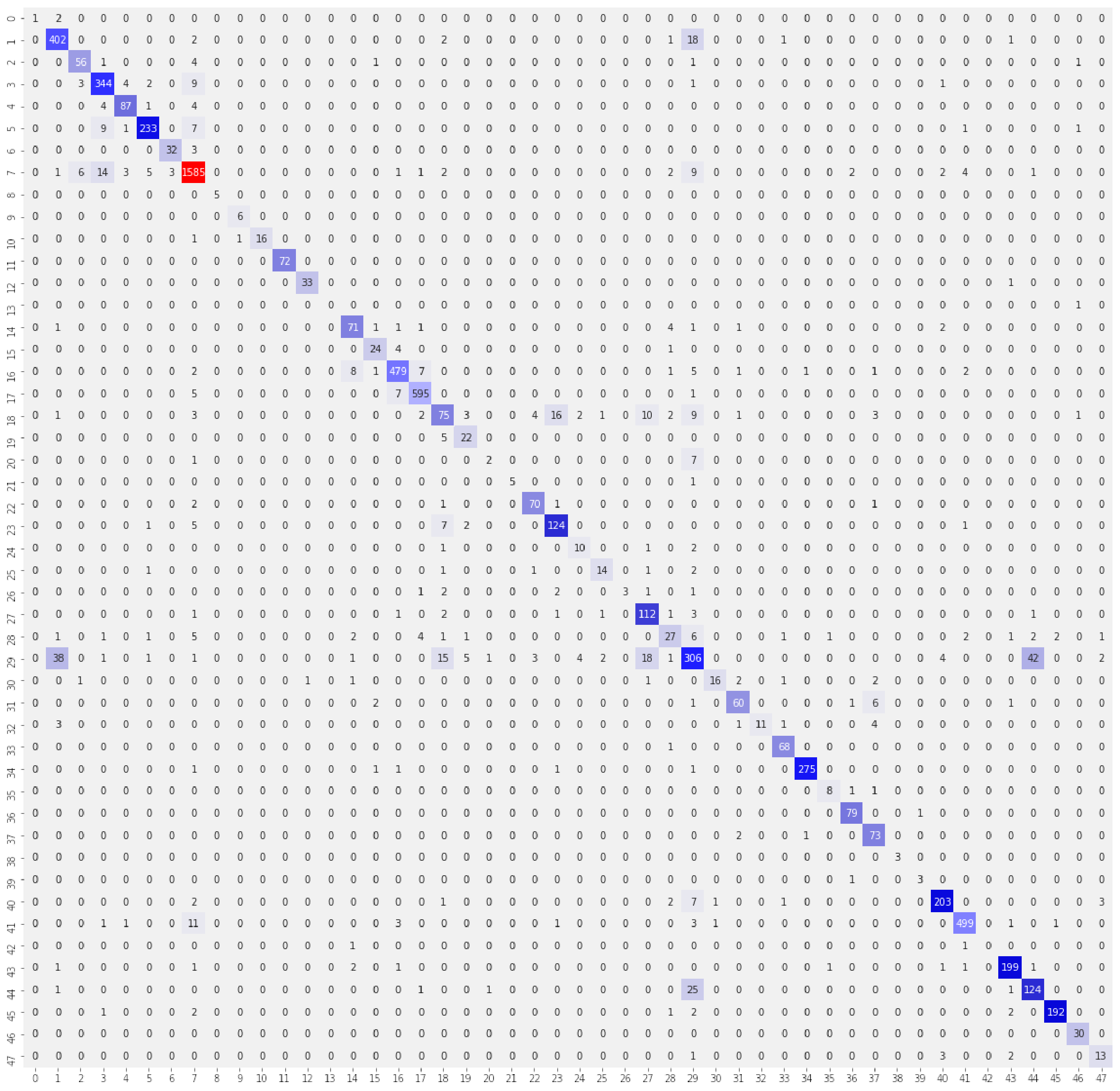}}}
\caption{The Confusion Matrices of Bert and RoBERTa Models}
\label{fig:conf_mat}
\end{figure*}

\begin{figure*}[ht]
\centering
\subfloat[Main Classes LSTM model]{\label{fig:LSTM_train}{\includegraphics[width=0.4\textwidth]{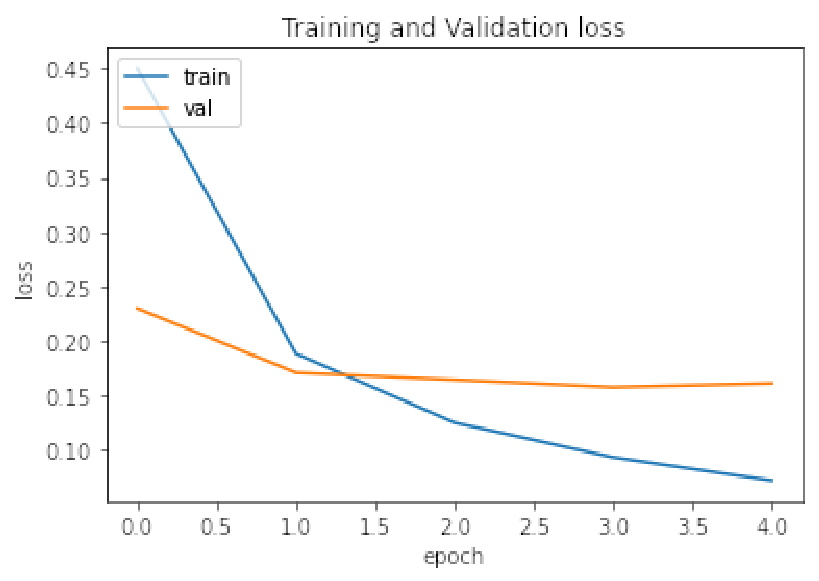}}}\hfill
\subfloat[Drugs Classes LSTM model]{\label{fig:LSTM_train_drugs}{\includegraphics[width=0.4\textwidth]{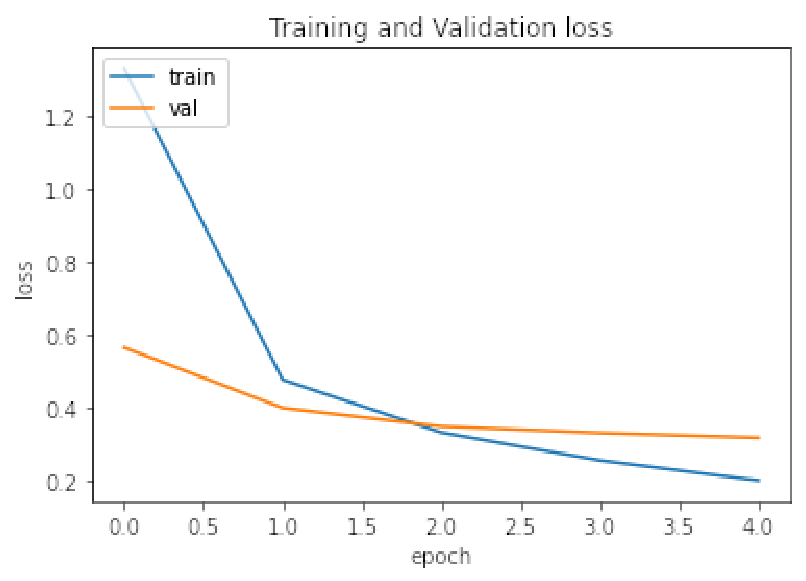}}}
\caption{Training and Validation Loss Plots per LSTM model}
\label{fig:subfiguresLSTM}
\end{figure*}

\begin{figure*}[ht]
\centering
\subfloat[First cycle]{\label{fig:ULMFIT_train1}{\includegraphics[width=0.3\textwidth]{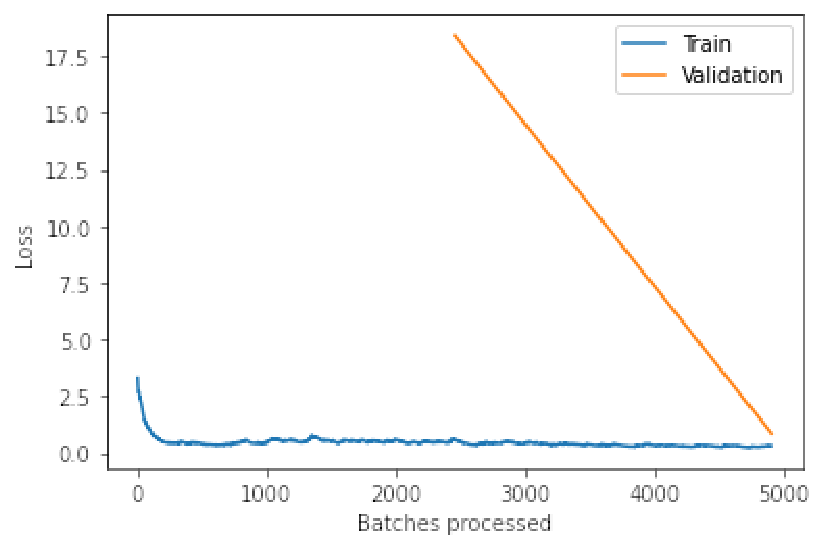}}}\hfill
\subfloat[Second cycle]{\label{fig:ULMFIT_train2}{\includegraphics[width=0.3\textwidth]{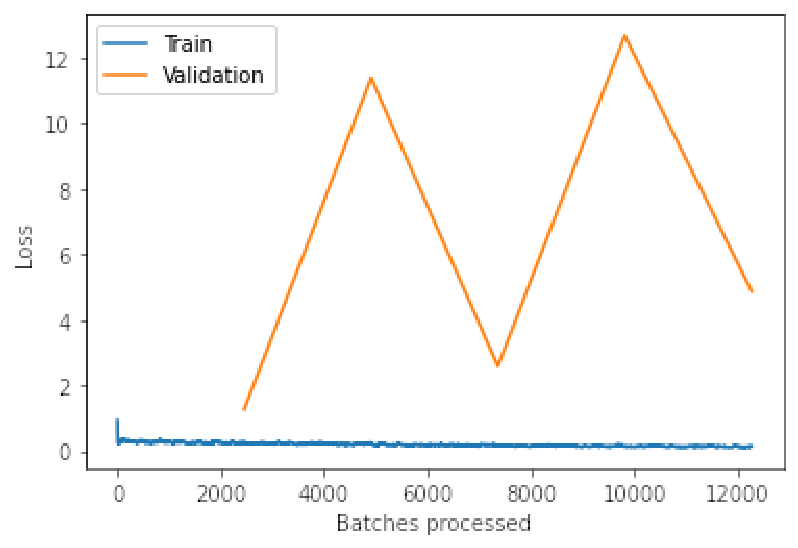}}}\hfill
\subfloat[Third cycle]{\label{fig:ULMFIT_train3}{\includegraphics[width=0.3\textwidth]{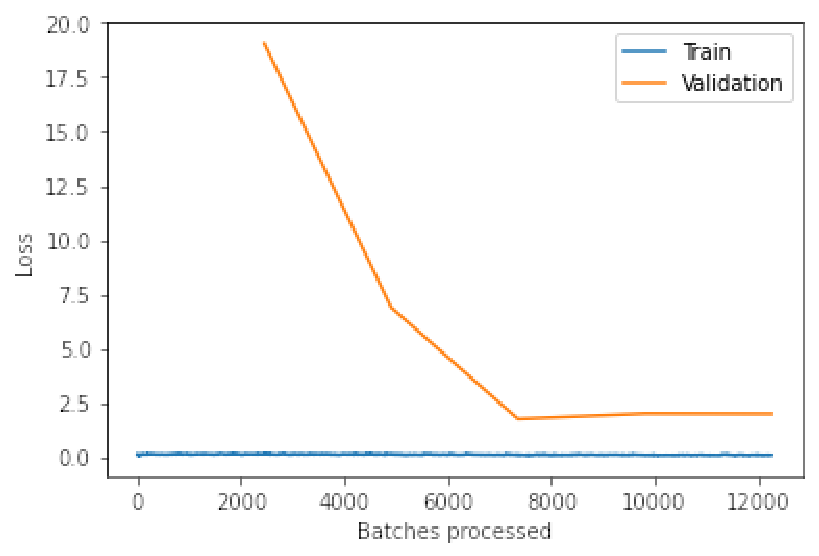}}}
\caption{Training and Validation Plots of the Main Classes ULMFIT model per learning cycle}
\label{fig:subfiguresULMFit}
\end{figure*}

\begin{figure*}[ht]
\centering
\subfloat[First cycle]{\label{fig:ULMFIT_drugstrain1}{\includegraphics[width=0.3\textwidth]{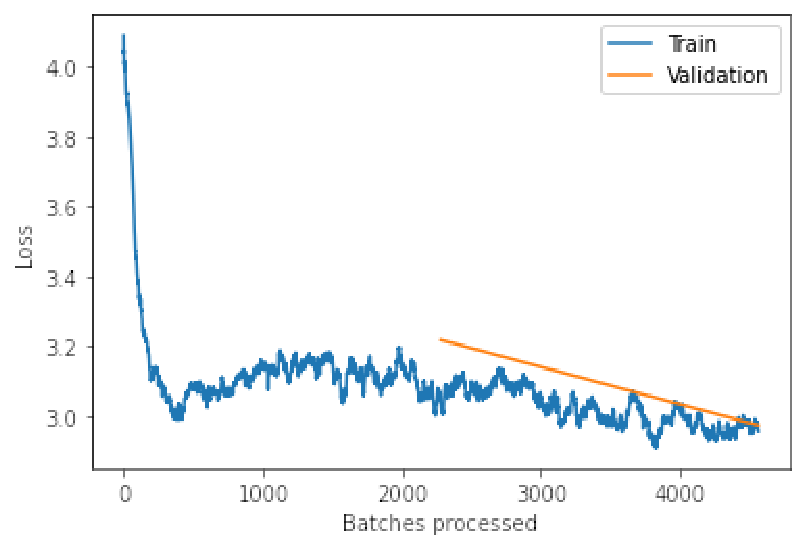}}}\hfill
\subfloat[Second cycle]{\label{fig:ULMFIT_drugstrain2}{\includegraphics[width=0.3\textwidth]{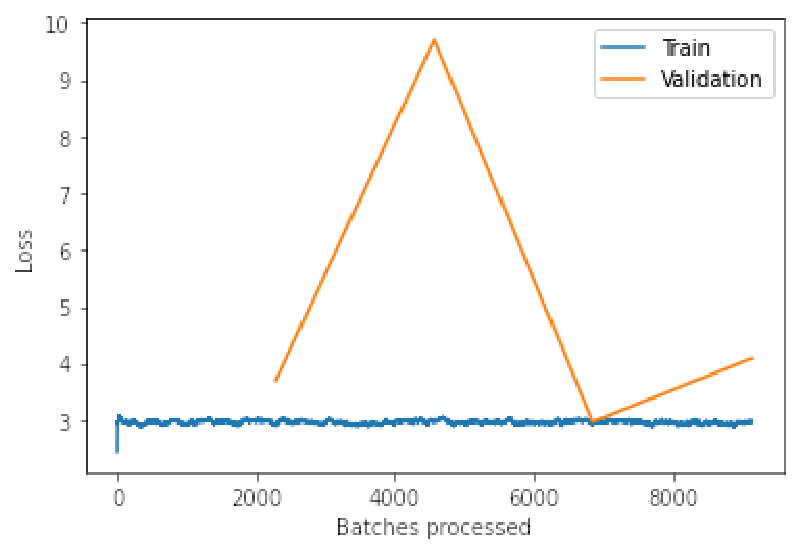}}}\hfill
\subfloat[Third cycle]{\label{fig:ULMFIT_drugstrain3}{\includegraphics[width=0.3\textwidth]{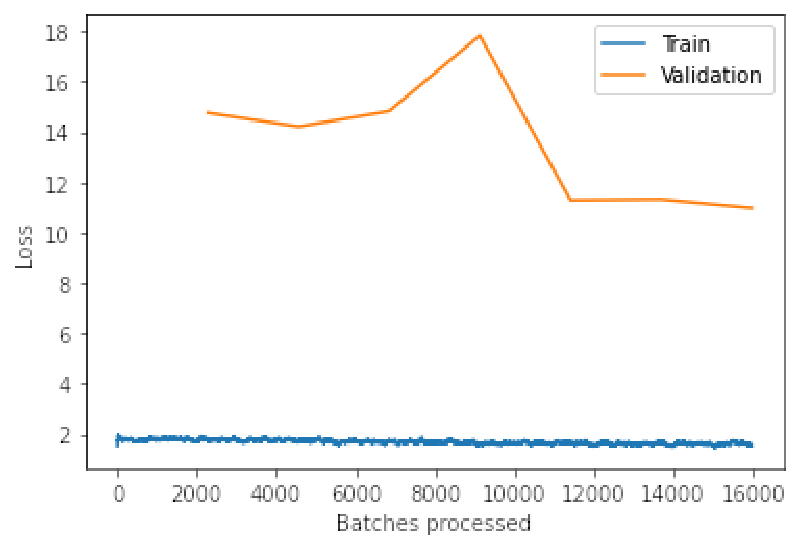}}}
\caption{Training and Validation Plots of the Drugs Classes ULMFIT model per learning cycle}
\label{fig:subfiguresULMFitdrugs}
\end{figure*}

\begin{figure*}[ht]
\subfloat[First cycle]{\label{fig:RoBerta_train1} {\includegraphics[width=0.3\textwidth]{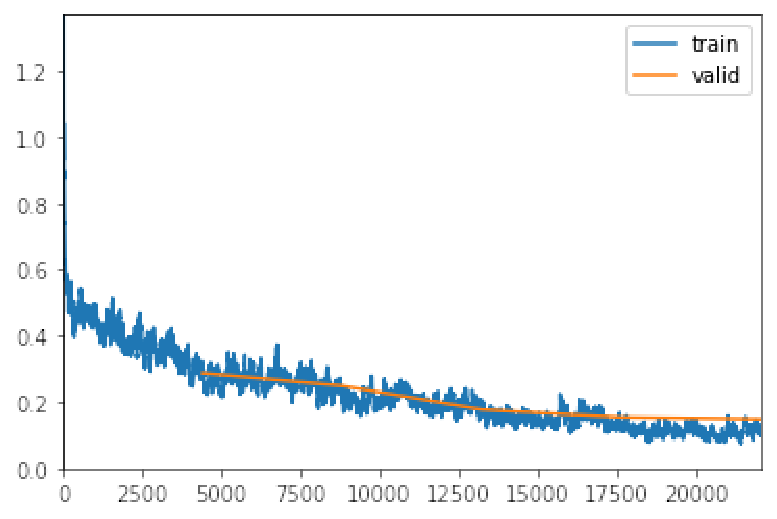}}}\hfill
\subfloat[Second cycle]{\label{fig:RoBerta_train2} {\includegraphics[width=0.3\textwidth]{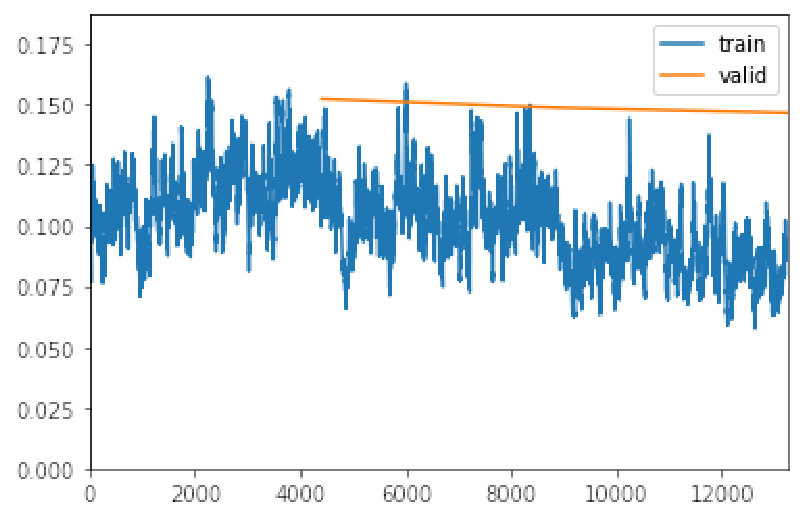}}}\hfill
\subfloat[Third cycle]{\label{fig:RoBerta_train3} {\includegraphics[width=0.3\textwidth]{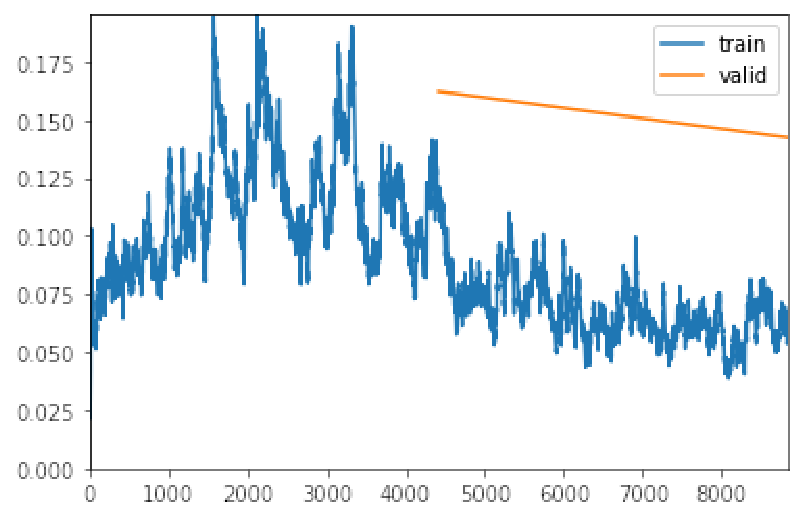}}}
\caption{Training and Validation Plots of the Main Classes RoBERTa model per learning cycle}
\label{fig:subfiguresRoBerta}
\end{figure*}

\begin{figure*}[ht]
\subfloat[First cycle]{\label{fig:RoBerta_drugstrain1} {\includegraphics[width=0.45\textwidth]{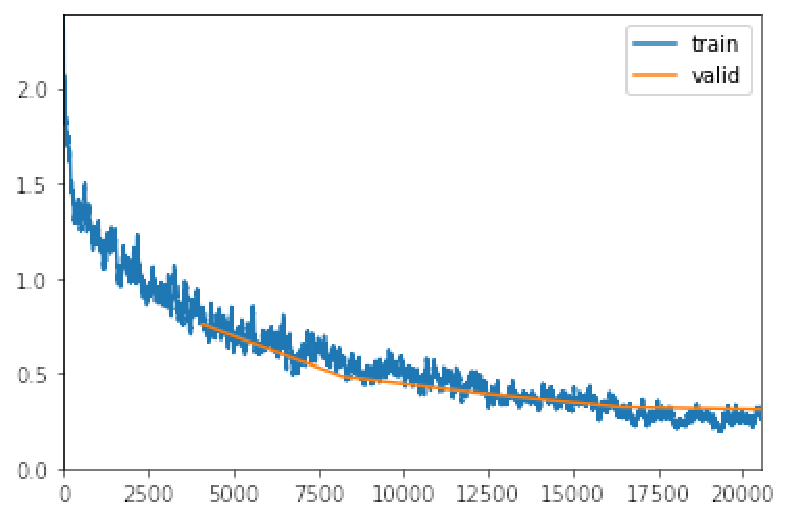}}}\hfill
\subfloat[Second cycle]{\label{fig:RoBerta_drugstrain2} {\includegraphics[width=0.45\textwidth]{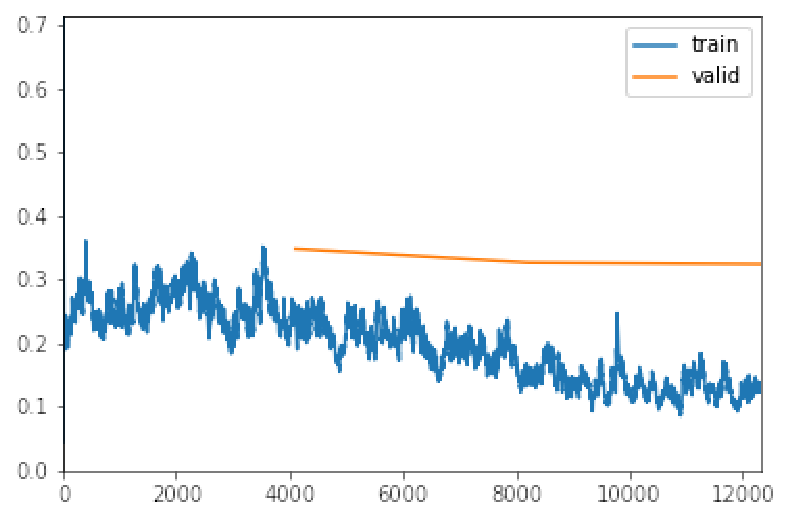}}}
\caption{Training and Validation Plots of the Drugs Classes RoBERTa model per learning cycle}
\label{fig:subfiguresRoBerta_drugs}
\end{figure*}

\end{document}